\documentclass[11pt]{article}
\usepackage[utf8]{inputenc}
\usepackage{graphicx}
\usepackage{epsfig}
\usepackage{latexsym}
\usepackage{amsmath, amsthm, amssymb, mathrsfs, amsfonts}
\usepackage{varioref}
\usepackage[table]{xcolor}
\usepackage{hhline}
\usepackage{mathtools}
\usepackage[inline]{enumitem}
\usepackage{longtable}
\usepackage[inline]{enumitem}
\usepackage{bbm}
\allowdisplaybreaks

\renewcommand{\theequation}{\arabic{equation}}
\newcommand{\beq}{\begin{equation}}
\newcommand{\eeq}{\end{equation}}

\newtheorem{theorem}{Theorem}

\newtheorem{lemma}{Lemma}
\newtheorem{corollary}{Corollary}

\newtheorem{definition}{Definition}

\newtheorem{example}{Example}

\def\ba{\begin{array}}
\def\ea{\end{array}}

\def\ni{\noindent}

\newcommand{\Prob}{\operatorname{P}}

\newcommand{\MI}{M\hspace{-1.0mm}I}
\newcommand{\BMI}{\mathbbm{I}}
\newcommand{\BSI}{\mathbbm{I}_{\scalebox{0.6}{S}}}

\newcommand\numberthis{\addtocounter{equation}{1}\tag{\theequation}}

\begin{document}
\pagestyle{myheadings} \markboth{Generalized Mutual Information}{Generalized Mutual Information}

\baselineskip24pt

\title{
Generalized Mutual Information  
  }
\author{Zhiyi Zhang
\\
University of North Carolina at Charlotte\\
Charlotte, NC 28223 }
\date{July 11, 2019}

\maketitle
\begin{abstract}
Mutual information is one of the essential building blocks of information theory. Yet, it is only finitely defined for distributions with fast decaying tails on a countable joint alphabet of two random elements. The unboundedness of mutual information over the general class of all distributions on a joint alphabet prevents its potential utility to be fully realized. This is in fact a void in the foundation of information theory that needs to be filled. This article proposes a family of generalized mutual information all of whose members 1) are finitely defined for each and every distribution of two random elements on a joint countable alphabet, except the one by Shannon, and 2) enjoy all utilities of a finite Shannon's mutual information.
\end{abstract}

\section{Introduction and Summary}   
Let $Z$ be a random element on a countable alphabet $\mathscr{Z}=\{z_{k};k\geq 1\}$ with an associated distribution $\mathbf{p}=\{p_{k};k\geq 1\}$. Let the cardinality or support on $\mathscr{Z}$ be denoted $K=\sum_{k\geq 1}1[p_{k}>0]$, where $1[\cdot]$ is the indicator function. $K$ is possibly finite or infinite. Let $\mathscr{P}$ denote the family of all distributions on $\mathscr{Z}$.
Let $(X,Y)$ be a pair of random elements on a joint countable alphabet $\mathscr{X}\times \mathscr{Y}=\{(x_{i},y_{j});i\geq 1, j\geq 1\}$ with an associated joint probability distribution $\mathbf{p}_{\scalebox{0.6}{X,Y}}=\{p_{i,j};i\geq 1,j\geq 1\}$, let the two marginal distributions be respectively denoted $\mathbf{p}_{\scalebox{0.6}{X}}=\{p_{i,\cdot}=\sum_{j\geq 1}p_{i,j};i\geq 1\}$ and $\mathbf{p}_{\scalebox{0.6}{Y}}=\{p_{\cdot,j}=\sum_{i\geq 1}p_{i,j};j\geq 1\}$. Let $\mathscr{P}_{X,Y}$ denote the family of all distributions on $\mathscr{X}\times \mathscr{Y}$.
 Shannon (1948) offers two fundamental building blocks of information theory, Shannon's entropy $H=H(Z)=-\sum_{k\geq 1}p_{k}\ln p_{k}$ and mutual information $\MI=\MI(X,Y)=H(X)+H(Y)-H(X,Y)$, where $H(X)$, $H(Y)$ and $H(X,Y)$ are entropies respectively defined with the distributions $\mathbf{p}_{\scalebox{0.6}{X}}$, $\mathbf{p}_{\scalebox{0.6}{Y}}$ and $\mathbf{p}_{\scalebox{0.6}{X,Y}}$. 

Mutual information plays a central role in the theory and the practice of modern data science for three basic reasons. First, the definition of $\MI$ does not rely on any metrization on an alphabet, nor does it require the letters of the alphabet to be ordinal. This generality allows it to be defined and used in data spaces beyond the real coordinate space $\mathbbm{R}^n$, where random variables (as opposed to random elements) reside. Second, when $X$ and $Y$ are random variables assuming real values, that is, the joint alphabet is metrized, $\MI(X,Y)$ captures linear as well as any non-linear stochastic association between $X$ and $Y$. See Chapter 5 of Zhang (2017) for examples. Third,
it offers a single-valued index measure for the stochastic association between two random elements, more specifically, $\MI(X,Y)\geq 0$ for any probability distribution of $X$ and $Y$ on a joint alphabet and $\MI(X,Y)=0$ if and only if $X$ and $Y$ are independent, under a wide class of general probability distributions.   

However mutual information $\MI$, in its current form, may not be finitely defined for joint distributions in a subclass of $\mathscr{P}_{X,Y}$, partially due to the fact that any or all of the three Shannon's entropies in the linear combination may be unbounded. 
The said unboundedness prevents the potential utility of mutual information to be fully realized, and hence is a deficiency of $\MI$ which leaves a void in $\mathscr{P}_{X,Y}$. (More detailed arguments are provided in Section 2 below.) This article introduces a family of generalized mutual information indexed by a positive integer $n\in \mathbb{N}$, denoted $\BMI=\{\MI_{n};n\geq 1\}$, each of whose members, $\MI_{n}$, is referred to as the $n^{th}$ order mutual information. All members of $\BMI$ are finitely defined for each and every  
$\mathbf{p}_{\scalebox{0.6}{X,Y}}\in \mathscr{P}_{X,Y}$, except $\MI_{1}=\MI$, and all of them preserve the utilities of Shannon's mutual information when it is finite. 

The said deficiency of $\MI$ is due to the fact that Shannon's entropy may not be finite for ``thick-tailed'' distributions (with $p_{k}$ decaying slowly in $k$) in $\mathscr{P}$. To address the deficiency of $\MI$, the issue of unboundedness of Shannon's entropy on a subset of $\mathscr{P}$ must be addressed, through some generalization in one way or the other. The effort to generalize Shannon's entropy has been long and extensive in the existing literature. The main perspective in the generalization in the existing literature is based on axiomatic characterization of Shannon's entropy. Interested readers may refer to Csisz\'{a} (2008) and Amig\'{o}, Balogh and  Hern\'{a}ndez (2018) for details and references therewithin. In a nut shell, with respect to the functional form, $H=\sum_{k\geq 1}h(p_{k})$, under certain desirable axioms, for example, Khinchin  (1957) and Chakrabarti and Chakrabarty (2005), $h(p)=-p\ln p$ is uniquely determined up to a multiplicative constant; if the strong additivity axiom is relaxed to be one of the weaker versions, say $\alpha$-additivity or composability, then $h(p)$ may be of other forms which give rise to R\'{e}nyi's entropy, by R\'{e}nyi (1961), and the Tsallis entropy, by Tsallis (1988). However all such generalization effort does not seem to lead to an information measure on a joint alphabet 
that would possess all the desirable properties of $\MI$, in particular $\MI(X,Y)=0$ if and only if $X$ and $Y$ are independent, which is supported by an argument via Kullback-Leibler divergence proposed by Kullback and Leibler (1951). 


Toward repairing the said deficiency of $\MI$, a new perspective of generalizing Shannon's entropy is introduced in this article. In the new perspective, instead of searching for alternative forms of $h(p)$ in $H=\sum_{k\geq 1}h(p_{k})$ under weaker axiomatic conditions, it is sought to apply Shannon's entropy to, not the original underlying distribution $\mathbf{p}$ but distributions induced by $\mathbf{p}$. One particular set of such induced distributions is a family, each of whose members is referred to as a conditional distribution of total collision (CDOTC) indexed by $n\in \mathbb{N}$. It is shown that Shannon's entropy defined with every  CDOTC induced by any $\mathbf{p}\in \mathscr{P}$ is bounded above, provided that $n\geq 2$. The boundedness of the generalized entropy allows mutual information to be defined for any CDOTC of degree $n\geq 2$ for any $\mathbf{p}_{\scalebox{0.6}{X,Y}}\in \mathscr{P}_{X,Y}$.
The resulting mutual information is referred to as the $n^{th}$ order mutual information index and is denoted $\MI_{n}$, which is shown to possess all the desired properties of $\MI$ but with boundedness guaranteed. The main results are given and established in Section 3 after several motivating arguments for the generalization of mutual information in Section 2. 

\section{Generalization Motivated} To further motivate the generalization of mutual information in this article, let the definition of mutual information be considered in a broader perspective. Inherited from the Kullback-Leibler divergence, mutual information on a joint alphabet, $\MI(X,Y)=\sum_{i\geq 1,j\geq 1}p_{i,j}\ln(p_{i,j}/(p_{i,\cdot}\times p_{\cdot,j}))$, is unbounded for a large subclass of distributions in $\mathscr{P}_{X,Y}$. 
Example \ref{example_infty_mi} below demonstrates the existence of such a subclass of joint distributions.
\begin{example} Let $\mathbf{p}=\{p_{k};k\geq 1\}$ be a probability distribution with $p_{k}>0$ for every $k$ but unbounded entropy. 
Let $\mathbf{p}_{\scalebox{0.6}{X,Y}}=\{p_{i,j}; i\geq 1\mbox{ and }j\geq 1\}$ be such that $p_{i,j}=p_{i}$ for all $i=j$ and $p_{i,j}=0$ for all $i\neq j$, hence 
$\mathbf{p}_{\scalebox{0.6}{X}}=\{p_{i,\cdot}=p_{i};i\geq 1\}$ and  $\mathbf{p}_{\scalebox{0.6}{Y}}=\{p_{\cdot,j}=p_{j};j\geq 1\}$. Then $\MI(X,Y)=\sum_{i\geq 1,j\geq 1}p_{i,j}\ln(p_{i,j}/(p_{i,\cdot}\times p_{\cdot,j}))=-\sum_{k\geq 1}p_{k}\ln p_{k}=\infty$. 
\label{example_infty_mi}
\end{example}

One of the most attractive properties of mutual information is that mutual information $\MI(X,Y)$ is finitely defined for all joint distributions such that $p_{i,j}=p_{i,\cdot}\times p_{\cdot,j}$ for all $i\geq 1$ and $j\geq 1$ and $\MI(X,Y)=0$ if and only if the two random elements $X$ and $Y$ are independent. However the utility of mutual information is beyond mere an indication of whether it is zero or not. The magnitude of mutual information is also of essential importance, although Shannon did not elaborate that in his landmark paper, Shannon (1948). The said importance is perhaps best illustrated by the notion of the standardized mutual information defined as $\kappa(X,Y)=\MI(X,Y)/H(X,Y)$ and Theorem \ref{KappaProp} below. However before stating Theorem \ref{KappaProp}, Definition \ref{121} below is needed.

\begin{definition} Random elements $X \in \mathscr{X}$ and $Y\in \mathscr{Y}$ are said to have an one-to-one correspondence, or to be one-to-one corresponded, under a joint probability distribution $\mathbf{p}_{\scalebox{0.6}{X,Y}}$ on 
  $\mathscr{X \times Y}$, if  
  \begin{enumerate}
    \item for every $i$ satisfying $\Prob(X=x_{i})>0$, there exists a unique $j$ such that $\Prob(Y=y_{j}|X=x_{i})=1$,
             and
    \item for every $j$ satisfying $\Prob(Y=y_{j})>0$, there exists a unique $i$ such that $\Prob(X=x_{i}|Y=y_{j})=1$.
  \end{enumerate}
\label{121}
\end{definition}  

\begin{theorem} Let $(X,Y)$ be a pair of random elements on alphabet $\mathscr{X}\times \mathscr{Y}$ with joint distribution $\mathbf{p}_{\scalebox{0.6}{X,Y}}\in \mathscr{P}_{X,Y}$ such that $H(X,Y)<\infty$. Then
  \begin{enumerate}
  \item  $0\leq \kappa(X,Y)\leq 1$,
  \item  $\kappa(X,Y)=0$ if and only if $X$ and $Y$ are independent, and 
  \item  $\kappa(X,Y)=1$ if and only if $X$ and $Y$ are one-to-one corresponded.
  \end{enumerate}
\label{KappaProp}
\end{theorem} A proof of Theorem \ref{KappaProp} can be found on page 159 of Zhang (2017). Theorem \ref{KappaProp} essentially maps independence of $X$ and $Y$ (the strongest form of unrelatedness) to $\kappa=0$, one-to-one correspondence (the strongest form of relatedness) to $\kappa=1$, and everything else in between. In so doing, the magnitude of mutual information is utilized in measuring the degree of dependence in pairs of random elements, which could lead to all sorts of practical tools in evaluating, ranking, and selecting variables in data space.  

It is important to note that the condition of $H(X,Y)<\infty$ is essential in Theorem \ref{KappaProp} since obviously, without it, $\kappa$ may not be well defined. In fact, if $H(X,Y)<\infty$ is not imposed, even observing reasonable conventions such as $1/\infty=0$ and $0/\infty=0$, the statements of Theorem \ref{KappaProp} may not be true. To see this, consider the following constructed example.

\begin{example}Let $\mathbf{p}=\{p_{k};k\geq 1\}$ be a probability distribution with $p_{k}>0$ for every $k$ but unbounded entropy. Let $\mathbf{p}_{\scalebox{0.6}{X,Y}}=\{p_{i,j}; \mbox{$i=1$ or 2 and }j\geq 1\}$ be such that 
\[
   p_{i,j}=\left\{\begin{array}{ll}
       p_{j}& \mbox{$i=1$ and $j$ is odd} \\
       p_{j}& \mbox{$i=2$ and $j$ is even} \\
       0& \mbox{otherwise,}
   \end{array}
   \right.
\]hence 
$\mathbf{p}_{\scalebox{0.6}{X}}=\{ p_{1,\cdot}, p_{2,\cdot} \}
=\{ \sum_{k=odd}p_{k},\sum_{k=even}p_{k}\}$ and  $\mathbf{p}_{\scalebox{0.6}{Y}}=\{p_{\cdot,j}=p_{j};j\geq 1\}$. $X$ and $Y$ are obviously not independent, and 
\begin{align*}
 0<\MI(X,Y)&=\sum_{i\geq 1,j\geq 1}p_{i,j}\ln(p_{i,j}/(p_{i,\cdot}\times p_{\cdot,j})) =H(X)<\infty.
  \end{align*} It follows that $\kappa=\MI(X,Y)/H(X,Y)=H(X)/H(X,Y)=0$ but in this case $\MI(X,Y)> 0$. Therefore Part 2 of Theorem \ref{121} fails.
\label{ExampleKappaZero}
\end{example}
Example \ref{ExampleKappaZero} indicates that mutual information in its current form is deprived of the potential utility of Theorem \ref{121} for a large class of joint distributions and therefore leaves much to be desired.

Another argument for the generalization of mutual information can be made in a statistical perspective. In practice, mutual information is often to be estimated from sample data. For statistical inference to be meaningful, the estimand $\MI(X,Y)$ needs to exist, {\it i.e.}, $\MI(X,Y)<\infty$. More specifically in testing the hypothesis of independence between $X$ and $Y$, $H_{0}: \mathbf{p}_{\scalebox{0.6}{X,Y}}\in \mathscr{P}_{0}$ where $\mathscr{P}_{0}\subset \mathscr{P}_{X,Y}$ is the subclass of all joint distributions for independent $X$ and $Y$ on $\mathscr{X}\times \mathscr{Y}$, $\MI(X,Y)$ needs to be finitely defined in an open neighborhood of $\mathscr{P}_{0}$ in $\mathscr{P}_{X,Y}$, or else the logic framework of statistical inference is not well supported. Let $\mathscr{P}_{\infty}$ be the subclass of $\mathscr{P}_{X,Y}$ such that $\MI(X,Y)=\infty$. In general, it can be shown that $\mathscr{P}_{\infty}$ is dense in $\mathscr{P}_{X,Y}$ with respect to the $p$-norm for $p\geq 1$. In specific, for any $\mathbf{p}_{\scalebox{0.6}{X,Y}}\in \mathscr{P}_{0}$, there exists a sequence of distributions $\{\mathbf{p}_{\scalebox{0.6}{m,X,Y}}\}\in \mathscr{P}_{\infty}$ such that $\|\mathbf{p}_{\scalebox{0.6}{m,X,Y}}- \mathbf{p}_{\scalebox{0.6}{X,Y}}\|_{p}\rightarrow 0$. See Example \ref{ExampleDense} below.
\begin{example}
Let $\mathbf{p}_{\scalebox{0.6}{X,Y}}=\{p_{i,j};\mbox{$i=1,2$ and $j=1,2$}\}$ where $p_{i,j}=0.25$ for all $(i,j)$ such that $1\leq i\leq 2$ and $1\leq j\leq 2$. Obviously $X$ and $Y$ are independent under $\mathbf{p}_{\scalebox{0.6}{X,Y}}$, that is, $\mathbf{p}_{\scalebox{0.6}{X,Y}}\in \mathscr{P}_{0}$. Let $\mathbf{p}_{\scalebox{0.6}{m,X,Y}}$ be constructed based on 
$\mathbf{p}_{\scalebox{0.6}{X,Y}}$ as follows. 

Remove an arbitrarily small quantity $\varepsilon/4>0$ where $\varepsilon=1/m$ away from each of the four positive probabilities in  $\mathbf{p}_{\scalebox{0.6}{X,Y}}$ so each becomes $p_{m,i,j}=0.25-\varepsilon/4$ for all $(i,j)$ such that $1\leq i\leq 2$ and $1\leq j\leq 2$. Extend the range of $(i,j)$ to $i\geq 3$ and $j\geq 3$, and allocate the mass $\varepsilon$ to over the extended range according to  
\[
    p_{m,i,j}=\left\{ \begin{array}{ll}
                      \dfrac{c}{i(\ln i)^2}& \mbox{$i\geq 3$, $j\geq 3$ and $i=j$} \\[8pt]
                      0& \mbox{$i\geq 3$, $j\geq 3$ and $i\neq j$}
                      \end{array}
              \right.
\] where $c$ is such that $\sum_{k\geq 3}c/[k(\ln k)^2]=\varepsilon$. Under the constructed $\{p_{m,i,j}\}$, for any $\varepsilon=1/m$, $X$ and $Y$ are not independent, and the corresponding mutual information is 
\begin{align*}
 \sum_{i\geq 1,j\geq 1} p_{m,i,j} \ln \left[
    \frac{p_{m,i,j}} { (p_{m,i,\cdot} p_{m,\cdot,j})}
    \right]&
  =4(0.25-\varepsilon/4)\ln\left[\frac{0.25-\varepsilon/4}{(0.5-\varepsilon/2)^2}\right]
    -\sum_{k\geq 3}\dfrac{c}{k(\ln k)^2}\ln \dfrac{c}{k(\ln k)^2}=\infty.
\end{align*} However noting, as $m\rightarrow \infty$, $\varepsilon\rightarrow 0$ and hence $c\rightarrow 0$,
\begin{align*}
\|\mathbf{p}_{\scalebox{0.6}{m,X,Y}}- \mathbf{p}_{\scalebox{0.6}{X,Y}}\|_{2}^2
 &=4\varepsilon^2+\sum_{k\geq 3}\left[\frac{c}{k(\ln k)^2}\right]^2=4\varepsilon^2+c^2\sum_{k\geq 3}\frac{1}{k^2(\ln k)^4} \rightarrow 0.
\end{align*}
\label{ExampleDense}
\end{example}

All things considered, it is therefore desirable to have a mutual information measure, say $\MI_{n}(X,Y)$, or for that matter a family of mutual information measures indexed by a positive integer $n$, such that $\MI_{n}(X,Y)<\infty$ for all distributions in $\mathscr{P}_{X,Y}$, and with an accordingly defined standardized mutual information measure $\kappa_{n}=\kappa_{n}(X,Y)$ such that the utility of Theorem \ref{121} is preserved with $\kappa_{n}$ in place of $\kappa$ for all distributions in $\mathscr{P}_{X,Y}$.

\section{Main Results}  Given $\mathscr{Z}=\{z_{k};k\geq 1\}$ and $\mathbf{p}=\{p_{k}\}$, consider the experiment of drawing an identically and independently distributed ($iid$) sample of size $n$. Let $C_{n}$ denote the event that all observations of the sample take on a same letter in $\mathscr{Z}$, and let $C_{n}$ be referred to as the event of total collision. The conditional probability, given $C_{n}$, that the total collision occurs at letter $z_{k}$ is
\beq 
p_{n,k}=\frac{p_{k}^n}{\sum_{i\geq 1}p_{i}^{n}}.
\label{pnk}  
\eeq It is clear that $\mathbf{p}_{n}=\{p_{n,k}\}$ is a probability distribution induced from 
$\mathbf{p}=\{p_{k}\}$.
\begin{lemma}For each $n$, $n\geq 1$, $\mathbf{p}$ and $\mathbf{p}_{n}$ uniquely determine each other. 
\label{lemma1}
\end{lemma}
\ni {\it Proof}.
Given $\mathbf{p}=\{p_{k};k\geq 1\}$, by (\ref{pnk}), $\mathbf{p}_{n}=\{p_{n,k};\geq 1\}$ is uniquely determined. Conversely, given $\mathbf{p}_{n}=\{p_{n,k};\geq 1\}$, for each $n$ and all $k\geq 1$,
$p_{k}^{n}/p_{1}^{n}=p_{n,k}/p_{n,1}$ and therefore 
\begin{align*}
 p_{k}&=p_{1}\left(\frac{p_{n,k}}{p_{n,1}} \right)^{1/n}, \hspace{1em}
 \sum_{i\geq 1}p_{i}=p_{1}\sum_{i\geq 1}\left(\frac{p_{n,i}}{p_{n,1}} \right)^{1/n}=1, \hspace{1em}
 p_{1}= \left[ \sum_{i\geq 1}\left(\frac{p_{n,i}}{p_{n,1}} \right)^{1/n}\right]^{-1}, \\
 p_{k}&=\left[ \sum_{i\geq 1}\left(\frac{p_{n,i}}{p_{n,1}} \right)^{1/n}\right]^{-1}\left(\frac{p_{n,k}}{p_{n,1}} \right)^{1/n}=\left[ \sum_{i\geq 1}\left(\frac{p_{n,i}}{p_{n,k}} \right)^{1/n}\right]^{-1}
 =\frac{ p_{n,k}^{1/n} }{  \sum_{i\geq 1} p_{n,i}^{1/n}}. \numberthis \label{pbypn}
\end{align*} The lemma follows.
\hfill $\Box$
\begin{lemma}
For each $n$, $n\geq 2$, and any $\mathbf{p}\in \mathscr{P}$, $H_{n}(Z)=-\sum_{k\geq 1}p_{n,k}\ln p_{n,k}<\infty$.
\label{lemma2}
\end{lemma}
\ni {\it Proof}. Write $\eta_{n}=\sum_{k\geq 1}p_{k}^{n}$. 
Noting $0<\eta_{n}\leq 1$ and $0\leq -p\ln p\leq 1/e$ for all $p\in [0,1]$,  
\begin{align*}
  H_{n}(Z)&=-\sum_{k\geq 1}p_{n,k}\ln p_{n,k}=-\sum_{k\geq 1}\frac{p_{k}^n}{\sum_{i\geq 1}p_{i}^{n}}\ln \frac{p_{k}^n}{\sum_{i\geq 1}p_{i}^{n}} \\
    &= -\frac{n}{\eta_{n}}\sum_{k\geq 1}p_{k}^{n}\ln p_{k} + \ln  \eta_{n}
    \leq \left(\frac{n}{e}\right)\left(\frac{\eta_{n-1}}{\eta_{n}}\right)+\ln \eta_{n}<\infty.
\end{align*} The lemma follows.
\hfill $\Box$

On the joint alphabet $\mathscr{X}\times \mathscr{Y}=\{(x_{i},y_{j})\}$ with distribution $\mathbf{p}_{\scalebox{0.6}{X,Y}}=\{p_{i,j}\}$, consider the associated CDOTC for an $n$ and all pairs $(i,j)$ such that $i\geq 1$ and $j\geq 1$,
\beq
    p_{n,i,j}=\frac{p_{i,j}^{n}}{\sum_{s\geq 1,t\geq 1}p_{s,t}^{n}}. 
\label{pnij}
\eeq Let $\mathbf{p}_{n,\scalebox{0.6}{X,Y}}=\{p_{n,i,j};i\geq 1,j\geq 1\}$. It is to be noted that $\mathbf{p}_{n,\scalebox{0.6}{X,Y}}\in \mathscr{P}_{\scalebox{0.6}{X,Y}}$. The two marginal distributions of (\ref{pnij}) are $\mathbf{p}_{n,\scalebox{0.6}{X}}
=\{p_{n,i,\cdot}\}$ and 
$\mathbf{p}_{n,\scalebox{0.6}{Y}}=\{p_{n,\cdot,j}\}$ respectively, where 
\begin{align*}
  p_{n,i,\cdot}&=\sum_{j\geq 1}p_{n,i,j}=\sum_{j\geq 1}\left(\frac{p^{n}_{i,j}}{\sum_{s\geq 1,t\geq 1}p^{n}_{s,t}}\right) = \frac{\sum_{j\geq 1}p^{n}_{i,j}}{\sum_{s\geq 1,t\geq 1}p^{n}_{s,t}},  
  \numberthis \label{pnx} \\
  p_{n,\cdot,j}&=\sum_{i\geq 1}p_{n,i,j}=\sum_{i\geq 1}\left(\frac{p^{n}_{i,j}}{\sum_{s\geq 1,t\geq 1}p^{n}_{s,t}}\right) = \frac{\sum_{i\geq 1}p^{n}_{i,j}}{\sum_{s\geq 1,t\geq 1}p^{n}_{s,t}}. 
  \numberthis \label{pny} 
\end{align*}
\begin{lemma}
 $\mathbf{p}_{\scalebox{0.6}{X,Y}}=\{p_{i,j}\}=\{p_{i,\cdot}\times p_{\cdot,j}\}$ if and only if 
 $\mathbf{p}_{n,\scalebox{0.6}{X,Y}}=\{p_{n,i,j}\}=\{p_{n,i,\cdot}\times p_{n,\cdot,j}\}$. 
\label{lemma3}
\end{lemma}
\ni {\it Proof}. For each positive integer $n$, if $p_{i,j}=p_{i,\cdot}\times p_{\cdot,j}$ for all pairs $(i,j)$, $i\geq 1$ and $j\geq 1$, then 
\begin{align*}
  p_{n,i,j}&= \frac{p_{i,j}^{n}}{\sum_{s\geq 1,t\geq 1}p_{s,t}^{n}}
  =\frac{p_{i,\cdot}^{n}p_{\cdot,j}^{n}}{\sum_{s\geq 1,t\geq 1}p_{s,\cdot}^{n}p_{\cdot,t}^{n}}
  =\left(\frac{p_{i,\cdot}^{n}}{\sum_{s\geq 1}p_{s,\cdot}^{n}}\right)
    \left( \frac{p_{\cdot,j}^{n}}{\sum_{t\geq 1}p_{\cdot,t}^{n}}\right)
\end{align*} and the two factors of the last expression above are respectively $\Prob(X_{1}=\cdots=X_{n}=x_{i}|C_{n})$ and $\Prob(Y_{1}=\cdots=Y_{n}=y_{j}|C_{n})$, $(X_{r},Y_{r})$, $r=1,\cdots, n$, are letter values of the $n$ observations in the sample. 

Conversely, if $p_{n,i,j}=p_{n,i}^* \times p_{n,j}^*$ where $p_{n,i}^*\geq 0$ depends only on $n$ and $i$ and $p_{n,j}^*\geq 0$ only depends on $n$ and $j$, then by (\ref{pbypn}),
\begin{align*}
 p_{i,j}&=\frac{ p_{n,i,j}^{1/n} }{  \sum_{s\geq 1,t\geq 1} p_{n,s,t}^{1/n}} 
 =\frac{ (p^*_{n,i})^{1/n}(p_{n,j}^*)^{1/n} }{  \sum_{s\geq 1} (p^*_{n,s})^{1/n} 
 \sum_{t\geq 1}(p^*_{n,t})^{1/n}} \\
 &=\left(
          \frac{ (p^*_{n,i})^{1/n} } {\sum_{s\geq 1}(p^*_{n,s})^{1/n}} 
   \right)
 \times 
   \left(
          \frac {(p_{n,j}^*)^{1/n} } {
 \sum_{t\geq 1}(p^*_{n,t})^{1/n}} \right).
\end{align*} The lemma immediately follows the factorization theorem.
\hfill $\Box$

For each $n\in \mathbb{N}$, let $H_{n}(X,Y)$, $H_{n}(X)$ and $H_{n}(Y)$ be Shannon's entropies defined with the joint CDOTC, $\{p_{n,i,j};i\geq 1\}$ as in (\ref{pnij}), and the  marginal distributions $\{p_{n,i,\cdot};i\geq 1\}$ and $\{p_{n,\cdot,j};j\geq 1\}$ as in (\ref{pnx}) and (\ref{pny}) respectively. 
Let 
\beq 
  \MI_{n}=\MI_{n}(X,Y)=H_{n}(X)+H_{n}(Y)-H_{n}(X,Y).
\label{MIn}
\eeq
\begin{theorem}
  For every $n\geq 2$ and any $\mathbf{p}_{\scalebox{0.6}{X,Y}}\in \mathscr{P}_{\scalebox{0.6}{X,Y}}$,
  \begin{enumerate}
  \item  $0\leq \MI_{n}(X,Y)<\infty$,
  \item  $\MI_{n}(X,Y)=0$ if and only $X$ and $Y$ are independent.
  \end{enumerate}
\label{theorem1}
\end{theorem}
\ni {\it Proof}.
In Part 1, $\MI_{n}\geq 0$ since $\MI_{n}$ is a mutual information and $\MI_{n}<\infty$ by Lemma \ref{lemma2}. Part 2 follows Lemma \ref{lemma3} and the fact that $\MI_{n}$ is a mutual information. 
\hfill $\Box$

Let 
\beq 
 \kappa_{n}=\kappa_{n}(X,Y)=\frac{H_{n}(X)+H_{n}(Y)-H_{n}(X,Y)}{H_{n}(X,Y)}
\label{kappan}
\eeq be referred to as the $n^{th}$ order standardized mutual information, and write $\BSI=\{\kappa_{n};n\geq 1\}$. 
Let $(X^*,Y^*)$ be a pair of random elements on $\mathscr{X}\times \mathscr{Y}$ according to the induced joint distribution $\mathbf{p}_{n,\scalebox{0.6}{X,Y}}$ with index value $n\geq 1$.
\begin{lemma}
\label{lemma4} $X$ and $Y$ have an one-to-one correspondence if and only if $X^*$ and $Y^*$ have one.
\end{lemma}
\ni {\it Proof}.
 If $X$ and $Y$ have an one-to-one correspondence, then for each $i$, there is a unique $j_{i}$ such that $p_{i,j_{i}}>0$ and $p_{i,j}=0$ for all other $j$, $j\neq j_{i}$. By (\ref{pnij}), $p_{n,i,j_{i}}>0$ and $p_{n,i,j}=0$ for all other $j$, $j\neq j_{i}$. That is, $X^*$ and $Y^*$ have an one-to-one correspondence. 
 
Conversely, if $X^*$ and $Y^*$ have an one-to-one correspondence, then for each $i$, there is a unique $j_{i}$ such that $p_{n,i,j_{i}}>0$ and $p_{n,i,j}=0$ for all other $j$, $j\neq j_{i}$. On the other hand, 
by (\ref{pbypn}),
\begin{align*}
p_{i,j}&
 =\frac{ p_{n,i,j}^{1/n} }{  \sum_{s\geq 1,t\geq 1} p_{n,s,t}^{1/n}},
 \end{align*}
it follows that $p_{i,j_{i}}>0$ and $p_{i,j}=0$ for all other $j$, $j\neq j_{i}$. That is, $X$ and $Y$ have an one-to-one correspondence. 
\hfill $\Box$
\begin{corollary}
For every $n\geq 2$ and any $\mathbf{p}_{\scalebox{0.6}{X,Y}}\in \mathscr{P}_{X,Y}$,
  \begin{enumerate}
  \item  $0\leq \kappa_{n}(X,Y)\leq 1$,
  \item  $\kappa_{n}(X,Y)=0$ if and only if $X$ and $Y$ are independent, and 
  \item  $\kappa_{n}(X,Y)=1$ if and only if $X$ and $Y$ are one-to-one corresponded.
  \end{enumerate}
\label{coro1}
\end{corollary}
\ni {\it Proof}. By Lemma \ref{lemma3}, $X$ and $Y$ are independent if and only if $X^*$ and $Y^*$ are. By Lemma \ref{lemma4}, $X$ and $Y$ are one-to-one corresponded if and only if $X^*$ and $Y^*$ are. The statement of Corollary \ref{coro1} follows directly from Theorem \ref{121}.
\hfill $\Box$

Theorem \ref{theorem1} and Corollary \ref{coro1} fill the void in $\mathscr{P}_{X,Y}$ left behind by $\MI$.

\end{document}